\documentclass[twocolumn,preprintnumbers,amsmath,amssymb]{revtex4}

\usepackage{graphicx}  
\usepackage{sidecap}
\begin{document}
\title{Negative terahertz conductivity at vertical carrier injection 
in\\
 a black-Arsenic-Phosphorus-Graphene  heterostructure integrated\\ with a light-emitting diode}
\author{Victor Ryzhii$^{1,2,3,4}$,  Maxim Ryzhii$^5$, Taiichi Otsuji$^1$, Valery  E.~Karasik$^4$, Vladimir~G.~Leiman$^5$, Vladimir Mitin$^6$,
and Michael  S. Shur$^7$}
\address{
$^1$Research Institute of Electrical Communication, Tohoku University, Sendai 980-8577, Japan\\
$^2$Institute of Ultra High Frequency Semiconductor Electronics of RAS,\\
 Moscow 117105, Russia\\
$^3$Center for Photonics and Infrared Technology, Bauman Moscow State Technical University, Moscow 111005, Russia\\
  $^4$  Center for
Photonics and Two-Dimensional  Materials, Moscow Institute of Physics Technology,
Dolgoprudny 141700, Russia\\
 $^5$Department of Computer Science and Engineering, University of Aizu, Aizu-Wakamatsu 965-8580, Japan\\
$^6$Department of Electrical Engineering, University at Buffalo, SUNY, Buffalo, New York, 
1460-1920 USA\\
$^7$ Department of Electrical, Computer, and Systems Engineering, Rensselaer Polytechnic Institute, Troy, New York 12180, USA\\
}
 \begin{abstract} 
{\bf Key words - Black arsenic-phosphorus, graphene, integration
with light-emitting diode, injection, carrier cooling, terahertz
lasing.}\\
We propose and analyze the heterostructure comprising  a black-arsenic-phosphorus layer (b-As$_{1-x}$P$_x$L) and a graphene layer (GL)
  integrated with a light-emitting diode (LED). The integrated b-As$_{1-x}$P$_x$L-GL-LED heterostructure can serve as an active part of the terahertz  (THz)  laser using the interband radiative transitions in the GL. The feasibility of the proposed concept is enabled by the combination of relatively narrow
  energy gap in the  b-As$_{1-x}$P$_x$L and the proper band alignment with the GL.
 The operation of the device in question is associated with the generation of the electron-hole pairs by
 the LED emitted near-infrared radiation in the b-As$_{1-x}$P$_x$L, cooling of the photogenerated electrons and holes in this layer, and their injection  into the GL. Since the minimum b-As$_{1-x}$PL energy gap 
 ($\Delta_G \simeq 0.15$~eV)
is smaller than  the energy of optical phonons in the GL, 
($\hbar\omega_0 \simeq 0.2$~eV),
the injection into the GL can lead to a 
relatively weak heating of the two-dimensional electron-hole plasma
(2D-EHP) in the GL. 
At the temperatures
somewhat lower than the room temperature, the injection can  cool  
the 2D-EHP. This is beneficial for 
the interband population inversion in the GL,
reinforcement of its negative dynamic conductivity, 
 and the realization of the optical and plasmonic modes lasing supporting the  new types of the THz radiation sources.
\end{abstract}

\maketitle

\section{Introduction}

The gapless energy spectrum of graphene layers (GLs)~\cite{1} supporting  the terahertz (THz) and far-infrared (FIR) radiative interband transitions
enables the detection, control, and generation of the THz and FIR radiation 
(see, for example, the review articles~\cite{2,3,4,5,6} and the references therein).
One of the most interesting potential application of the GLs and the GL-based heterostructures is their use
in efficient THz and FIR lasers that were predicted to operate at room temperature and already demonstrated operation at 100 K~\cite{7,8,9,10,11,12,13,14,15,16,17,18,19,20}. Such lasers
can be particularly useful in the spectral range below 5 to 10 THz where the operation of the heterostructure lasers based on A$_3$B$_5$
compounds is hampered by a strong radiation absorption by the optical phonons.
Both the optical and injection pumping of  GL-based heterostructures can lead to the interband population inversion and enable the negative dynamic conductivity in the THz and FIR spectral ranges.
The quantum efficiency of the optical pumping into the GLs is limited   by relatively low absorption coefficient of the GLs ($\beta = \pi\alpha \simeq 0.023$, where $\alpha = e^2/c\hbar \simeq 1/137$ is the fine structure constant,
$e$ is the electron charge, $\hbar$ is the reduced Planck constant, and $c$ is the speed of light in vacuum). 
Apart from this, the generation of the electron-hole pairs in the GL by the near- and mid-infrared (NIR and MIR) photons,
i.e., by the photons with relatively high energy can lead to a substantial
 heating of the two-dimensional electron-hole plasma (2D-EHP) in the GL. This is because,
the initial energy, $\varepsilon_0$, of the electrons and holes generated by the incident
NIR photons ($\varepsilon_0 = \hbar\Omega/2$, where $\hbar\Omega \sim 1$~eV is the photon energy of the pumping radiation). 
As a result, the effective temperature of the 2D-EHP could be rather large.
This  complicates the achievement of the interband population inversion in the GLs~\cite{22}.  
The low absorption  limitation can be partially avoided in the heterostructures including multiple non-Bernal stacked GLs~\cite{9}. 
The  quantum efficiency of the optical pumping can be enhanced, for example, by using a bulk absorption layer, in which the incident NIR/MIR radiation with the photon energy $\hbar\Omega$ exceeding the energy gap of this layer $\Delta_G$ generates the electron-hole pairs followed by  their vertical diffusion into the GL~\cite{21} (perpendicular to the GL plane). 
However, in the case 
of the GaAs (or similar materials) as the absorption layer~\cite{21}, although  the pumping quantum efficiency 
can be markedly increased, the injected electrons and holes are still fairly hot, because the carriers are injected into the GL still with a high energy $\varepsilon_i \gtrsim \Delta_G \sim 1$~eV.  
In principle, the GL optical pumping by \lq\lq warm \rq\rq carriers can be realized using the CO$_2$ or mid-infrared (MIR)
quantum cascade lasers. But  the optical pumping by NIR/MIR light-emitting diodes (LEDs) or lasers appears to be much more practical.  

In this paper, we propose to use for the GL-based THz and FIR lasers the absorption layer with the sufficiently narrow energy gap and the proper band alignment with the GL~\cite{23,24,25,26,27}.
 As for the material for such a layer can be choosen black-arsenic-phosphorus (b-As$_{1-x}$P$_x$) or black-arsenic (b-As). The  b-As$_{1-x}$P$_x$ layer comprising a relatively large number of the atomic sheets can exhibit the energy gap $\Delta_g \simeq 0.15$~eV when the phosphorus fraction $x$ is small~\cite{28,29,30,31} (at $x = 0.17$, i.e., close to b-As). Further decrease in $x$, i.e.,
 up to pure b-As may push the band gap to even smaller values~\cite{28}. In the case of the b-As$_{1-x}$P$_x$ absorbing layer,
 the optical pumping can be provided by a source of NIR/MIR radiation as previously, but due to an effective cooling of the generated and propagated carriers in this layer, their energy of the injected pair  being about $\Delta_G$ can be much smaller than $\hbar\Omega$. 
 Thus, the absorption layer with relatively narrow energy gap
 can play an extra role of the carrier cooler. Such a combination can enable both the realization of a high pumping quantum efficiency and the injection of "warm" carriers. 
 As show in the paper, the injection into the GL
 of the carriers with the  energy smaller than the energy of optical phonons in the GL $\hbar\omega_0 \simeq 0.2$~eV can result in the 2D-EHP strong cooling. 
 The latter reinforces the interband population inversion
in the GL and the effect of its negative dynamic conductivity.
Recent advances in the b-As$_{1-x}$P$_x$-based heterostructure fabrication (see, for example,~\cite{29,30,31,32}
are in favor of the feasibility of the proposed lasers. 

\begin{figure}[t]
\centering
\includegraphics[width=8.0cm]{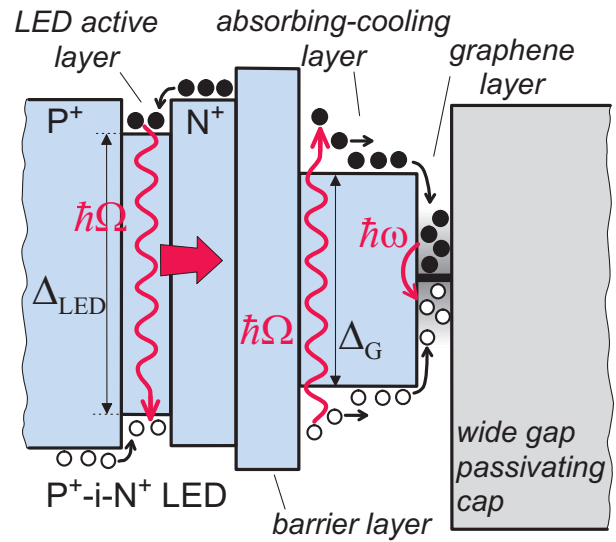}
\caption{Schematic view of the band diagram of a b-As$_{1-x}$P$_{x}$-GL  heterostructure integrated with a P$^+$-i-N$^+$- LED (at sufficiently strong forward bias).}
\label{F1}
\end{figure}

\section{Device structure and operation principle}

Figure~1 shows the schematic  band diagram of the device under consideration.
The device comprises   the b-As$_{1-x}$P$_x$L -GL heterostructure playing the role of the THz active region (THz-AR)   mounted on the top of the P$^+$-i-N$^+$ heterostructure serving as an LED. The THz-AR and LED are separated by a wide-gap transparent barrier layer (TBL). The NIR/MIR radiation (with the photon energy $\hbar\Omega \gtrsim \Delta_{LED}$) generated by the LED passes the TBL and produces the electrons and holes in the  b-As$_{1-x}$P$_x$L (with the initial kinetic energy  $(\varepsilon_0 = \hbar\Omega- \Delta_G)/2$. Since the energy gap, $\Delta_{LED}$
 is markedly larger that the energy gap  $\Delta_G$, in the b-As$_{1-x}$P$_x$L, the photogenerated
 carriers in the latter are fairly hot $\varepsilon_0 \gg k_BT_0$, where $k_B$ and $T_0$ are the Boltzmann constant and the ambient lattice temperature). If the thickness, $d$, of the   b-As$_{1-x}$P$_x$L substantially exceeds the characteristic cooling length $l_{\varepsilon}$ (the carrier energy relaxation length),
the photogenerated carriers arrive at the GL being efficiently cooled down the energy $3k_BT_0/2$.
Hence, the electron-hole pairs photogenerated in  b-As$_{1-x}$P$_x$L are injected into the GL having the net energy 
$\varepsilon_i = \Delta_G + 3k_BT_0$. If $\Delta_{LED} \sim 1$~eV, $\varepsilon_i \ll \Delta_{LED}$.
 
The cap layer is intended to preserve the GL or it can be a part of THz waveguide. 
The GL on the device top can be covered by a polycrystalline or by an organic polymer dielectric layer~\cite{33}.
The top layer  can be also made of different materials, for example, from  hexagonal boron nitride (hBN). This material can provide the enhanced dynamic properties of the carriers in the GL beneficial for achieving the negative THz conductivity. 
However, the interaction of the carriers with the interfacial 
optical phonons can lead to a complex pattern of the interband and intraband relaxation processes in the GLS~\cite{34,35}.

Below we consider the  laser heterostructure  with the GL active region pumped by 
 the injection of the electron-hole pairs  into the absorption-cooling  the b-As$_{1-x}$P$_x$ layer (b-As$_{1-x}$P$_x$L) with a small
phosphorous content $x$

It is assumed that the thickness, $d$, and  the absorption coefficient, $\beta_{\Omega}$, of the NIR/MIR radiation  with the energy $\hbar\Omega > \Delta_G$ in  b-As$_{1-x}$P$_x$L satisfy the following condition:

\begin{equation}\label{eq1} 
\beta_{\Omega}^{-1}, l_{\varepsilon}  \ll d < l_D.
\end{equation}
 Here  $l_{\varepsilon}$ is the characteristic length of the carrier energy relaxation (cooling) and $l_D$ is the carrier ambipolar diffusion across the absorption layer. Since  the energy relaxation time of the carriers photoexcited
in the absorption layer  $\tau_{\varepsilon} =\tau_{\varepsilon}|_{\varepsilon = (\hbar\Omega - \Delta_G)/2}$ can be  assumed to be much shorter than the recombination time $\tau_R$, the ratio $l_{\varepsilon}/l_D \simeq \sqrt{\tau_{\varepsilon}/\tau_R} \ll 1$. For the photon energy $\hbar\Omega$, the absorption coefficient can be set as $\beta_{\Omega} \sim 10^{5}$~cm$^{-1}$. Hence, for $d \gtrsim 1~\mu$m,
inequality~(1) should be valid.
 
According to inequality~(1), the generation of the electron-hole pairs by the pumping radiation
and their cooling occur primarily close to the irradiated surface of the absorption layer ($z = 0$, the axis $z$
is directed perpendicular to the absorption layer and the GL plane).
Therefore, the electron-hole density $n$ in the latter layer obeys the  diffusion equation:

\begin{equation}\label{eq2} 
-D_0\frac{d^2 n}{d z^2} + \frac{n}{\tau_R} = \beta_{\Omega}I_{\Omega}\exp(-\beta_{\Omega}z).
\end{equation}
The boundary conditions for Eq.~(2) are as follows:

\begin{equation}\label{eq3} 
D_0\frac{d n}{d z}\biggl|_{z = 0} = 0,\qquad -D\frac{d n}{d z}\biggl|_{z = d} = C_{GL}n|_{z=d},
\end{equation}
Here  $D_0$ and $\tau_R$ are the coefficient of the carrier ambipolar diffusion 
perpendicular to
the absorption layer and their recombination time 
 at the lattice temperature $T_0$, respectively, $I_{\Omega}$ is
  the photon flux incident on the absorption layer, and $C_{GL}$ is the rate of the carrier capture into the GL (the capture velocity~\cite{36,37}). Introducing the external quantum efficiency of the LED $\eta_{\Omega}$
 one can express $I_{\Omega}$ via the electric  power $P_{LED}$ consuming by the LED: $I_{\Omega} = \eta_{\Omega} P_{LED}/A\hbar\Omega$, where $A$ is the area of the LED and GL. 
 
 The flux of the electrons and holes, $J$,  injected into the GL (captured by the GL) and the flux of the energy, $Q$,  brought by
 the injected carriers to the 2D-EHP are equal to, respectively,

\begin{equation}\label{eq4} 
J = -D_0\frac{d n}{d z}\biggl|_{z = d}, \qquad Q = -\varepsilon_iD_0\frac{d n}{d z}\biggl|_{z = d},
\end{equation}
 where $\varepsilon_i = \Delta_G + 3k_BT_0$ and $k_B$ is the Boltzmann constant.

Solving Eq.(2) with boundary conditions (3) and using Eq.~(4), we obtain

\begin{equation}\label{eq5} 
J = \frac{I_{\Omega} C_{GL}}{({D_0}/l_D)\sinh(d/l_D) + C_{GL}\cosh(d/l_D)}, 
\end{equation}

\begin{equation}\label{eq6} 
 Q =\varepsilon_i J,
\end{equation}
where $l_D =\sqrt{D\tau_R}|_{T= T_0}$ is the carrier  ambipolar diffusion length.
Equations~(5) and (6) are valid if the carriers are photogenerated and cooled down in a narrow layer adjacent to the absorption layer surface (as assumed in line with left-side of inequality (1).

Since the carrier density in the GL under the laser operation conditions should be sufficiently large
(to provide the 2D-EHP degeneration), the inter-carrier collisions, characterized by a short  carrier scattering time $\tau_{cc}$ could lead to a "Fermitization" of
the distribution functions with the common effective temperature $T_e = T_h =  T$. Hence, the latter can be presented as $f_e(\varepsilon_h) = [\exp(\varepsilon_e -\mu_e)/k_BT +1]^{-1}$ and $f_h(\varepsilon_h) = [\exp(\varepsilon -\mu_h)/K_BT +1]^{-1}$, where $\varepsilon_e,\varepsilon_h, \mu_e$, and $\mu_h$ are the electron and hole energies and the electron and hole quasi-Fermi
energies counted from the Dirac point in the GL.

The quantity $C_{GL}$ depends on the electron and hole densities in the GL. 
The capture of the photogenerated carriers propagating across the absorption layer is accompanied  by the emission of optical phonons and inter-carrier scattering. 
An increase in the 2D-EHP density in the GL and, hence, an increase in the quasi-Fermi energies $\mu_e$ + $\mu_h$, leads to
a decrease of the capture. To account for this effect, we set 

\begin{equation}\label{eq7} 
C_{GL} = C_0\Theta\biggl(\frac{\Delta_G  - \mu_e -\mu_h}{\Delta_G}\biggr).
\end{equation}
Here $C_0$ is the capture velocity into the empty GL and $\Theta(\delta) = \delta\cdot \theta(\delta)$, where
$\theta(\delta$ is the Heaviside step function ($\theta(\delta) = 1$ if $\delta \geq 0$ and $\theta(\delta) = 0$ if $\delta < 0$.
For the heterostructures with the same electron and hole parameters ($\Delta_C = \Delta_V = \Delta_G/2$),
considered in the following, one can put $\mu_e = \mu_h = \mu$.
In this case, from Eqs.~(5) and (7) we obtain

\begin{equation}\label{eq8} 
J = I_{\Omega}\,\frac{ \Theta\displaystyle\biggl(\frac{\Delta_G  - 2\mu}{\Delta_G}\biggr)}
{\biggl[\displaystyle S_0\sinh\biggl(\frac{d}{l_D}\biggr) + \displaystyle \cosh\biggl(\frac{d}{l_D}\biggr)\Theta\biggl(\frac{\Delta_G  - 2\mu}{\Delta_G}\biggr)\biggr]},
\end{equation} 
where
$S_0 = D_0/C_0l_D$ is the parameters characterizing the carrier capture into the GL.

\section{Balance equations}

\begin{figure*}[t]
\centering
\includegraphics[width=15.0cm]{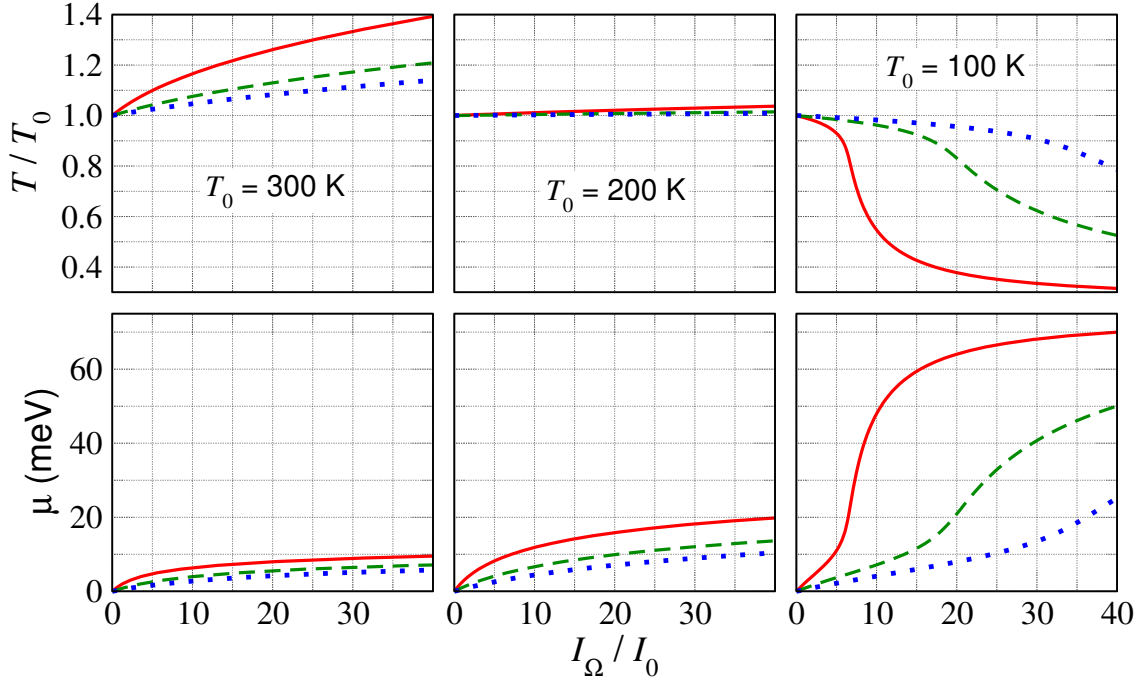}
\caption{Normalized effective temperature $T/T_0$ and quasi-Fermi energy $\mu$ versus $I_{\Omega}/I_0$ calculated for
different values of parameter $S_0$ ($S_0 = 1$ -solid lines, $S_0=5$ -dashed lines, and $S_0=10$ - dotted lines) and lattice temperatures: $T_0 = 300$~K,   $T_0 = 200$~K, and  $T_0 = 100$~K.
 }
\label{F2}
\end{figure*}

\begin{figure*}[t]
\centering
\includegraphics[width=15.0cm]{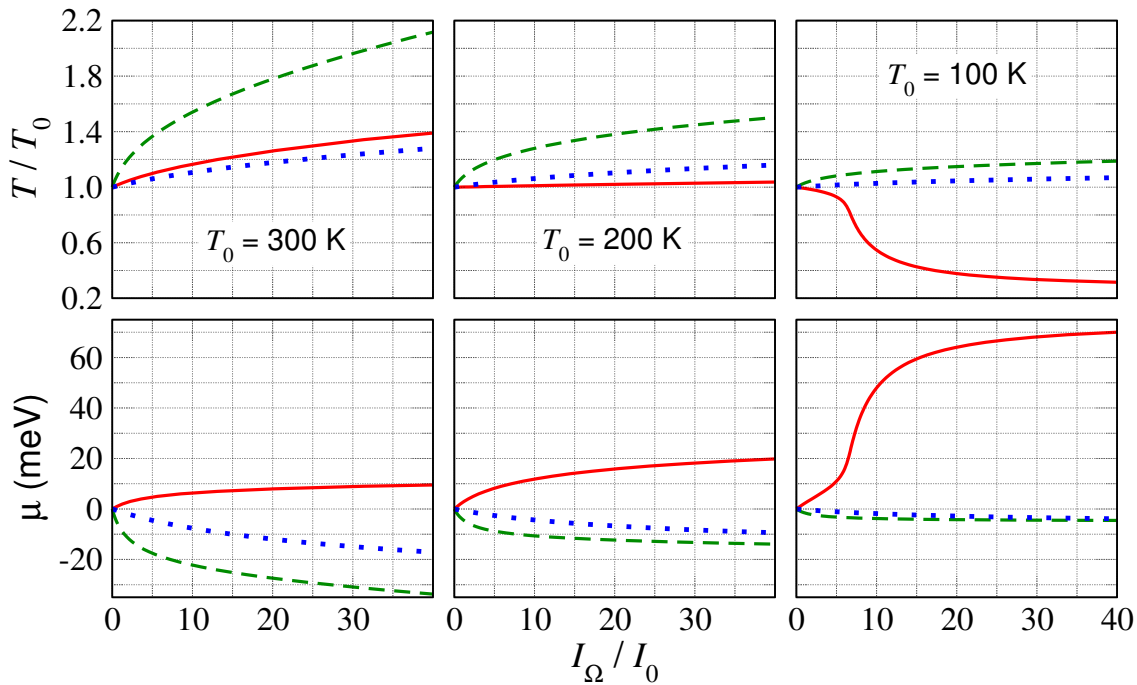}
\caption{Normalized effective temperature $T/T_0$ and quasi-Fermi energy $\mu$ as functions of $I_{\Omega}/I_0$
 for the devices with a narrow-gap absorbing-cooling layer (b-AsL) - solid lines, with a wide-absorbing layer (b-PL) - dashed lines, and with direct optical pumping of the GL ($\hbar\Omega = 0.36$~meV)
- dotted lines at $T_0 = 300$~K $T_0 = 200$~K, and $T_0 = 100$~K (for $S_0 = 1$).
 }
\label{F3}
\end{figure*}

The intersubband and intraband carrier relaxation in the GLs under pumping is primarily determined by the interaction with the GL optical phonons~\cite{38} (see also,~\cite{7,9,10,22}.
The direct Auger processes in the GLs are virtually prohibited~\cite{39,40} due to the  linear   carrier energy spectra~\cite{1}. Even though more complex Auger processes can 
contribute to the interband carrier balance~\cite{39,40,41}, we
 disregard these processes.
  Considering that in each act of the interband and intraband  emission/absorption  of the GL optical phonons 
 the 2D-EHP energy 
decreases/increases by the quantity  $ \hbar\omega_0$,  
 we present the equation governing the energy balance as~\cite{11,22}

\begin{eqnarray}\label{eq9}
\exp\biggl(\frac{2\mu}{k_BT} \biggl)
\exp\biggl[\hbar\omega_0k_B\biggl(\frac{1}{T_0} - \frac{1}{T}\biggr) \biggr] -1\nonumber\\
+a\biggl\{\exp\biggl[\hbar\omega_0k_B\biggl(\frac{1}{T_0} - \frac{1}{T}\biggr)\biggr] -1\biggr\}
 = 
 \frac{J}{I_0}\biggl(\frac{\varepsilon_i}{\hbar\omega_0} \biggr).
\end{eqnarray}
The interband balance of  electrons and holes is described by

\begin{eqnarray}\label{eq10}
\exp\biggl(\frac{2\mu}{k_BT} \biggl)
\exp\biggl[\hbar\omega_0k_B\biggl(\frac{1}{T_0} - \frac{1}{T}\biggr) \biggr] - 1
 = \frac{J}{I_0}.
\end{eqnarray}
Here $I_0 = \Sigma_0/\tau_{Opt}^{inter}$
is the rate of the electron-hole pairs generation due to the absorption of the thermal equilibrium optical phonons with the lattice temperature $T_0$,  $\Sigma_0$ is the characteristic carrier density determined by the energy dependence of the density of state in the GL near the Dirac point,    $a =\tau_{Opt}^{inter}/\tau_{Opt}^{intra}$
is the ratio of the pertinent times characterizing the interband transitions, 
$\tau_{Opt}^{inter}$ and $\tau_{Opt}^{intra}$ are the characteristic recombination and intraband relaxation
times associated with the carrier interaction with the optical phonons ($\tau_{Opt}^{inter} < \tau_{Opt}^{intra} $~\cite{22}) with $\tau_{Opt}^{intra} \sim \tau_0\exp(\hbar\omega_0/k_BT_0)$  being
  larger than the characteristic  time  of the optical phonon spontaneous emission $\tau_0$ (which is shorter than  1~ps) by a large factor of  $\exp(\hbar\omega_0/k_BT_0)$.
At $T_0 = 300$K and $T_0 = 100$~K, one can set~\cite{22,38} $I_0 \sim 10^{21}$~cm$^{-2}$s$^{-1}$
and $I_0 \sim 10^{14}$~cm$^{-2}$s$^{-1}$.

 The left-hand sides of Eqs.~(9) and (10) correspond to  the processes of the interband and intraband energy relaxation and the recombination-generation processes. The right-hand sides of these equations represent   the energy  and carrier fluxes into the GL normalized by $I_{0}$.
 
 \section{Carrier effective temperature and  quasi-Fermi energies versus pumping}
 
Equations~(9) and (10) yield
 
\begin{eqnarray}\label{eq11}
T = \frac{T_0} {1 - \displaystyle\frac{T_0}{\hbar\omega_0}\ln \biggl[1 +\frac{J}{I_0}\biggl(\frac{\Delta_G +3k_BT_0}{\hbar\omega_0} - 1\biggr)\frac{1}{a}\biggr]},
\end{eqnarray}

\begin{eqnarray}\label{eq12}
\frac{\mu}{T} = 
\frac{1}{2}\ln\biggl[\frac{1 +  \displaystyle\displaystyle\displaystyle\frac{J}{I}}
{1  +\displaystyle\frac{J}{I_0}\biggl(\frac{\Delta_G + 3k_BT_0}{\hbar\omega_0} - 1\biggr)\frac{1}{a}}\biggr].
\end{eqnarray}
 Equation~(11) and (12) together with Eq.~(8) describe the variation of the carrier effective temperature
 and  quasi-Fermi energy.
 
In particular, at the lattice temperature $T_0 = T_0^* = (\hbar\omega_0 - \Delta_G)/3 \simeq 0.05$~eV
($T_0^* \simeq 200$~K), Eqs.~(11) and (12) yield $T = T_0* = const$

 In line with the experimental data related to the b-P~\cite{42} for the 
carrier
mobility in the direction perpendicular to the atomic sheets in the b-As$_{1-x}$P$_x$ at the temperatures about the room temperature one can set $b \sim 400$~cm$^2$/V$\cdot$s. This gives $D_0 \sim 10$~cm$^2$/s. Considering that the carrier lifetime
in similar material is about $\tau_R \gtrsim 400$~ps~\cite{43}, we find $l_D \gtrsim 0.6~\mu$m. The capture velocity
can be estimated in the range~\cite{36,37} $C_0 \sim 10^4 - 10^5$~cm/s. As a result, assuming that $d \simeq l_D \simeq 1\mu$m, one can find
$S_0 \simeq 1 - 10$. 
 
 Figure~2 shows the effective temperature $T$ normalized by the lattice temperature $T_0$ and the quasi-Fermi energy $\mu$ calculated using Eqs.~(8), (11), and (12)
 as functions of the normalized pumping intensity $I_{\Omega}/I_0 \propto P_{LED}$ for $d = l_D$ and different values of parameter 
 $S_0 = D_0/C_0l_D = 1 - 10$
 in  the lattice temperatures range   $T_0= 100 - 300$~K . It is assumed that $\Delta_G = 0.150$~eV and
$\hbar\omega_0 = 0.2$~eV. 

As seen from Fig.~2, the $T/T_0$ versus $I_{\Omega}/I_0$ dependences for different lattice temperatures
are qualitatively different: at $T_0 = 300$~K, $T_0 = 200$~K, and $T_0 = 100$~K these dependences are rising (the heating of the 2D-EHP), constant (the effective temperature virtually does not change),
and decreasing (the cooling of the 2D-EHP), respectively. 
These distinctions are associated with
 the difference in the ratio of the power injected by the carriers into the GL and removed by the optical phonons, determined by the quantity $\varepsilon_i/\hbar\omega_0$.
Indeed, at $T_0 = 300$~K, $\varepsilon_i = 0.225$~eV, i.e., $\varepsilon_i > \hbar\omega_0$.
On the contrary, at $T_0 = 100$~K,   $\varepsilon_i = 0.175$~eV  corresponding  to $\varepsilon_i < \hbar\omega_0$.  While in the first case, the injection of the electron-hole pair into the GL 
increases the 2D-EHP energy by the value $\varepsilon_i - \hbar\omega_0 \simeq 0.025$~eV,
in the second case the 2D-EHP energy decreases (by the value $\hbar\omega_0 - \varepsilon_i \simeq 0.025$~eV.
When $T_0 = T_0* = 200$~K, the effective temperature is constant coinciding with $T_0$.

At all  lattice temperatures under consideration, the quasi-Fermi energy $\mu$ increases with increasing
$I_{\Omega}/I_0$. This increase, being moderate at $T_0 = 300$~K, becomes very steep at $T_0 = 100$~K, so that the pumping efficiency
 $\propto \mu/I_{\Omega}$ could be large.

One needs to point out that the plots related to the different lattice temperatures
correspond fairly different values of $I_{\Omega}$ because of a strong dependence of $I_{0}$ on $T_0$
(see the above estimates for $I_0$). This, in particular, implies that at $T_0 = 100$~K a significant change
in $T/T_0$ and $\mu$ can be achieved at the photon flux $I_{\Omega}$ much smaller (by several orders of magnitude)  than at $T_0 = 300$~K. Thus, a decrease of $T_0$ is an important factor for the enhancement
of the pumping efficiency.

Figure~3 compares the pumping efficiency of  the proposed laser heterostructure
with the absorption-cooling narrow-gap layer (b-AsL, $x =0$) and of the device having a wide-gap absorbing layer (b-PL, $x = 1$ and $\Delta_G = 0.3$~eV) - both with $S_0 = 1$, as well   the GL-based heterostructure with the direct optical pumping of the GL without the absorbing layer (see Appendix A).
We assumed that all the devices under comparison are irradiated by a MIR laser such as having the InAs active region, 
setting $\hbar\Omega = 0.36$~meV. The coefficient of the pumping  radiation absorption in the GL is $\beta = 0.023$. 
As seen from Fig.~3, 
in the  case of the absence of the preliminary carrier cooling in the absorbing layer and $\Delta_G = 0.3$~eV , a strong increase in the effective temperature leads to $\mu <0$, i.e., to the 2D-EHP nondegeneracy. The same occurs in the case of the direct pumping with $\hbar\Omega = 0.36$~meV.
The suppression of the population inversion despite an increase in the carrier density in the GL is, in this particular case,  associated with an excessive rise
of the effective temperature.

In contrast, the loss of the  carrier energy  in the absorption-cooling layer can lead to
 a  moderate increase  in $T$ with increasing $I_{\Omega}$ at $T_0 = 300$~K, to
an insensitivity of  $T$ to $I_{\Omega}$ at $T_0 =200$K, and even to a marked drop of $T$
at $T_0 = 100$~K. The latter implies that  pumping results in the 2D-EHP cooling.
 In all these cases, $\mu$ exhibits an increase, which is particularly steep at $T_0 = 100$~K,
 corresponding to a strong 2D-EHP degeneracy and population inversion.

 When the energy of the pumping photons $\hbar\Omega > 2\hbar\omega_0$,
the situation can be different, i.e., the population inversion in
the devices without carrier cooling in the absorption layer and in the devices with the direct optical pumping can be also achieved.
At such photon energies, that the carriers injected from the absorbing layer with $\Delta_G \lesssim \hbar\Omega$ into the GL
and the carriers directly photogenerated in the GL might emit a cascade of the optical phonons before the carrier Fermitization. 
This requires that  the time, $\tau_0$, of the spontaneous emission of the optical phonons in the GL should be sufficiently short in comparison  with the inter-carrier collision time $\tau_{cc}$.
At the pumping by the high energy photons, the mutual collisions of the photogenerated carriers having relatively high energy
can be characterized by not too short $\tau_{cc}$. 
In this case, the effective initial energy of the electron-hole pair in the GL
can be estimated as $\varepsilon_i^{eff} = \hbar\Omega - 2K\hbar\omega_0/(1 + K\tau_0/\tau_{cc})
$\,~\cite{22} (see also Appendix A),
where $K$ is the number of the optical phonons which can be emitted by the carrier injected into the GL. 
Considering this situation,  
from Eqs.~(A3) and (A4) one can find that the following three cases can be realized:\\
(a) $
a < \displaystyle\frac{\Omega}{\omega_0} - 1 - \frac{2K}{1 + K(\tau_0/\tau_{cc})}$, corresponding to  $d T/d I_{\Omega} > 0$
(carrier heating) and $d \mu/d I_{\Omega} < 0 $ ($\mu < 0$, hence, no population inversion);\\
(b) $
0 < \displaystyle\frac{\Omega}{\omega_0} - 1 - \frac{2K}{1 + K(\tau_0/\tau_{cc})} < a$, corresponding to 
$d T/d I_{\Omega} > 0$
(carrier heating) and $d \mu/d I_{\Omega} > 0 $ ($\mu > 0$,  population inversion);\\
(c) $\displaystyle\frac{\Omega}{\omega_0} - 1 - \frac{2K}{1 + K(\tau_0/\tau_{cc})} < 0$, corresponding to 
$d T/d I_{\Omega} < 0$ (carrier cooling) and $d \mu/d I_{\Omega} > 0 $ ($\mu > 0$,  population inversion).

The  population inversion in the 2D-EHP accompanied by its the  cooling occurs, for example,
in the case of direct optical  pumping by a CO$_2$ laser ($\hbar\Omega \sim 0.1$~eV, so that the latter inequality is satisfied ($K = 0$ and $\Omega/\omega_0 - 1 \simeq - 0.5$). In particular, if $\Omega/\omega_0 = 2 +\delta$ with $\delta < 1$, $K = 1$, and the condition of both cooling and population inversion at the direct optical pumping can be presented as $\tau_0/\tau_{cc}< (1 - \delta)/(1 + \delta)$.

\section{Dynamic conductivity and THz amplification}

The contributions of the direct interband optical transitions to the real part of the 2D-EHP dynamic conductivity 
Re$\sigma_{\omega}^{inter}$ can be found as in the previous papers
~\cite{9,10,11,12} (see also Refs.~\cite{7,45,46,47}): 

\begin{equation}\label{eq13}
{\rm Re} \sigma_{\omega}^{inter} \simeq \frac{e^2}{4\hbar}\tanh\biggl(\frac{\hbar\omega - 2\mu}{4k_BT}\biggl).
\end{equation}

One can see that in the range $\hbar\omega < 2\mu$,  Re$\sigma_{\omega}^{inter} < 0$. This implies that
the 2D-EHP can serve as an active region for the amplification of the  photonic or surface plasmon
modes propagating along the GL and their lasing.
However, the spectral range where Re$\sigma_{\omega}^{inter} < 0$ is limited
by not too large $\hbar\omega > \hbar\omega_D \sim \hbar/\tau$ by the carrier intraband absorption (the Drude absorption).
In reality, the value $\hbar\omega_D$, which is primarily determined by the carrier momentum relaxation
time $\tau_p$  ($\omega_D \sim D/2\pi\tau$)
in the GL, can be much smaller than $2\mu$. At $\tau_p \sim 1- 3$~ps, the absolute value of the dynamic conductivity for $\hbar\omega \sim 10 - 30$~meV can be only slightly smaller that $e^2/4\hbar$.
In this case, the surface plasmon amplification coefficient can be fairly large (about $\alpha_{\omega} \sim 10^4$~cm$^{-1}$). 

The carriers in the absorbing-cooling layer, particularly, in a narrow vicinity of the GL can lead  to an increase in the "parasitic" absorption of the THz radiation emitted 
by the laser heterostructure at the population inversion in the GL. Therefore, the carrier density
in this region
$n_{GL} \simeq n|_{z = d -0}$ should be limited.
This density   can be estimates as
 
 \begin{eqnarray}\label{eq14}
 n_{GL} = \frac{J}{C_{GL}} \nonumber\\
 = 
\frac{n_0}{\biggl[\displaystyle S_0\sinh\biggl(\frac{d}{l_D}\biggr) + \displaystyle \cosh\biggl(\frac{d}{l_D}\biggr)\Theta\biggl(\frac{\Delta_G  - 2\mu}{\Delta_G}\biggr)\biggr]}\,
\frac{I_{\Omega}}{I_0},
\end{eqnarray}
where $n_0 = I_0/C_0$. At $T_0 = 300$~K setting $C_0 = (10^4 - 10^5)$~cm/s, one obtains $n_0 \sim (10^{16} - 10^{17})$~cm$^{-3}$.

Figure 4 shows the normalized carrier density $n_{GL}/n_0$  as a function of the normalized pumping intensity $I_{\Omega}/I_0$ calculated for  $T_0 = 300$~K  and different values of $S_0$ (other parameters are the same as for Fig.~2).

These values of the absorption coefficient mentioned above, are markedly higher than the absorption coefficient, $-\alpha_{\omega}^{a-c}$, associated with
the free carrier  absorption in the absorbing-cooling layer near the GL where the plasmons are located. 
(in the direction perpendicular to the GL plane)~\cite{35}. At $n_{GL} \simeq n_0 \simeq (10^{16} - 10^{18})$, i.e., at the values corresponding to Fig.~4,
one obtains $\alpha_{\omega}^{*} \simeq 3\times(10 - 1000)$~cm$^{-1}$, i.e.,  $\alpha_{\omega}^{*}
\ll \alpha_{\omega}$.

\begin{figure}[t]
\centering
\includegraphics[width=7.0cm]{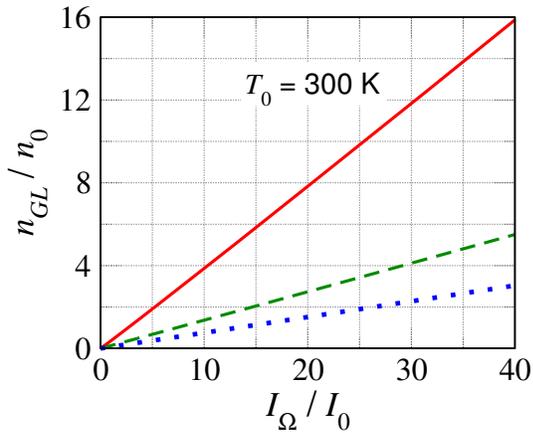}
\caption{Normalized carrier density in the absorption layer $n_{GL}/n_0$ versus $I_{\Omega}/I_0$ at $T_0 = 300$~K calculated for
and different values of $S_0$: $S_0 = 1$ -solid line, $S=5$ -dashed line, and $S_0=10$ - dotted line.
 }
\label{F4}
\end{figure}

 \section{Discussion}

Apart from this,
The recombination and the intraband energy relaxation lead to the generation of nonequilibrium
(hot) optical phonons in the GL (heating of the optical phonon system).
This system  cools down through anharmonic decay to acoustic phonons which are subsequently absorbed into the substrate~\cite{33,48,49,50}.
However, it was shown
 experimentally,  that the optical phonon decay time in the GL-heterostructures, in particular,
 the GL-hBN heterostructures  is estimated to be about~\cite{33} $\tau_{Opt}^{decay} \sim 0.200 - 0.375$~ps. At such  decay times, the deviation of the optical phonon system from equilibrium is insignificant.  This implies that this system temperature is close to the equilibrium temperature
 $T_0$.  In particular, in the case of the heterostructures with, in particular,  
 the hBN top layer, due to the large specific heat capacity of this layer, the rise of the lattice temperature even under relatively strong pumping is small ($\sim 1$~K)~\cite{33}.

The  carriers generated in the absorption-cooling layer transfer their excessive energy to the lattice of this layer. The heating of this layer can lead to an increase in $T_0$ if the heat drain is insufficient.
From the above results we can see that at lowered lattice temperatures, a sufficiently strong population 
 inversion (a large ratio $\mu/T$ can be achieved at moderate photon fluxes of the pumping radiation and, hence, using not so powerful LEDs. In this case, the device heating might be weak.
At elevated lattice temperatures, the heat power generated in the device can be substantial despite
high heat conductivity of the materials in the device in question (in particular, the GLs).
This might necessitate  using  special device configurations promoting an effective  heat removal of
work in the pulse regime. The devices withe absorption-cooling layer proposed and  considered by us
can demonstrate higher pumping efficiency in comparison   that with the absorption layer~\cite{21}.
However, both  devices 
 exhibit the same lattice heating provided the same pumping photon energy and flux. The point is that
 the excessive photon energy converts into the lattice heat in the absorbing-cooling layer in the former
 device, while this energy goes  directly to the 2D-EHP (resulting in its stronger heating) in the GL in the latter.

The possibility to enhance the pumping efficiency of in the  heterostructure under consideration
is associated with a relatively low energy gap in the absorbing-cooling layer and the proper
band alignment of this layer material and the GLs. In principle, other narrow-gap materials can be used.

We considered the device in which the pumping source (LED) is integrated into the device structure.
Naturally, the GL-based heterostructures with b-As$_{1-x}$P$_x$L or similar absorbing-cooling layer
with the optical pumping by separate sources can be used for the THz lasing.

To realize a practical laser device with the proposed heterostructure a pertinent laser cavity with good mode-field confinement needs to be integrated. The THz photon generation due to
the  transitions in the GLs includes both the in-plane and vertical modes. As is similar to a vertical cavity surface emitting laser diode, a Fabry-Perot vertical cavity might be  implemented along with the substrate (with a back-side full-reflection mirror coat and a top-side high reflection mirror coat).  Also analogously to  an edge-mode laser diode a distributed feedback cavity as well as a distributed brag reflector cavity can be implemented.

\section*{Conclusion}

We propose to use a  b-As$_{1-x}$P$_x$L-GL heterostructure
integrated with a NIR/MIR LED as an active region for the terahertz lasers.
the b-As$_{1-x}$P$_x$L sandwiched between the LED and GL serves for the LED radiation absorption, the carrier  generation and  cooling
followed by their injection. Using the b-As$_{1-x}$P$_x$Ls, exhibiting narrow energy gaps, enables
the injection of relatively cool carriers. This provides a higher pumping
efficiency in comparison with the laser GL-based heterostructures with relatively wide-gap  absorbing layer
and the laser GL-based heterostructures with the direct optical pumping.
We demonstrated that in the devices with p-AsL ($x \simeq 0$), i.e., with the absorbing-cooling layer characterized by the energy gap smaller than
the GL-optical phonon energy,  the carrier effective temperature in the GL can become lower than the lattice temperature. This might result in a further enhancement of the pumping efficiency
and the THz laser performance.

The authors are grateful to V. Ya. Aleshkin and A. A. Dubinov for useful data related to the surface plasmon absorption by the free carriers in the absorbing-cooling layer.

The work at RIEC and UoA was supported by Japan Society for Promotion of Science (Grants Nos.
16H-06361 and 16K-14243), the works at MIPT and RPI were supported by
Russian Foundation for Basic Research (Grant Nos. 18-29-02089 and 18-07-01379) and by Office of Naval Research (Project Monitor Dr. Paul Maki), respectively.

\section*{Appendix A.Direct optical pumping }
\setcounter{equation}{0}
\renewcommand{\theequation} {A\arabic{equation}}

In the case of the direct pumping of the GL by the LED NIR/MIR radiation, Eqs. (5), (9), and (10)
should be replaced by the following~\cite{22}:

\begin{eqnarray}\label{eqA1}
\exp\biggl(\frac{2\mu}{k_BT} \biggl)
\exp\biggl[\hbar\omega_0k_B\biggl(\frac{1}{T_0} - \frac{1}{T}\biggr) \biggr] -1\nonumber\\
+a\biggl\{\exp\biggl[\hbar\omega_0k_B\biggl(\frac{1}{T_0} - \frac{1}{T}\biggr)\biggr] -1\biggr\}
 = 
 \beta\frac{I_{\Omega}}{I_0}\biggl(\frac{\varepsilon_i^{eff}}{\hbar\omega_0} \biggr),
\end{eqnarray}

\begin{eqnarray}\label{eqA2}
\exp\biggl(\frac{2\mu}{k_BT} \biggl)
\exp\biggl[\hbar\omega_0k_B\biggl(\frac{1}{T_0} - \frac{1}{T}\biggr) \biggr] - 1
 = \beta\frac{I_{\Omega}}{I_0}.
\end{eqnarray}
Here $\beta = 0.023$, $\varepsilon_i^{eff} = \hbar\Omega - b\hbar\omega_0$, where
$b = 2K/(1 + K\tau_0/\tau_{cc})$, $K$ is the number of optical phonons in their cascade in the GL,
and $\tau_0$ and $\tau_{cc}$ were defined above.

Solving Eqs.~(A1)
 and (A2), we obtain
 
\begin{eqnarray}\label{eqA3}
\frac{T}{T_0} =\frac{1}{1 - \displaystyle\frac{k_BT_0}{\hbar\omega_0}
\ln\biggl[1 + \frac{\beta}{a}\frac{I_{\Omega}}{I_0}\biggl(\frac{\Omega}{\omega_0} - b- 1\biggr)\biggr]}\nonumber\\
\simeq 
\biggl(\frac{k_BT_0}{\hbar\omega_0}\biggr)\biggl(\frac{\beta}{a}\biggr)\biggl(\frac{\Omega}{\omega_0} - b - 1\biggr)\,\frac{I_{\Omega}}{I_0},
\end{eqnarray}

\begin{eqnarray}\label{eqA4}
\frac{\mu}{T} =\frac{1}{2}\ln\biggl[\frac{1 + \displaystyle\beta\frac{I_{\Omega}}{I_0}}{1 + \displaystyle\frac{\beta}{a}\biggl(\frac{\Omega}{\omega_0} - b - 1\biggr)}\frac{I_{\Omega}}{I_0}\biggr]\nonumber\\
\simeq \frac{\beta}{2}\biggl[1 - \frac{1}{a}\biggl(\frac{\Omega}{\omega_0} - b - 1\biggr)\,\frac{I_{\Omega}}{I_0}.
\end{eqnarray}

\end{document}